\documentclass[aps,
reprint,
amsmath, 
amssymb,
nofootinbib,
raggedbottom,
superscriptaddress]{revtex4-2}
\usepackage{graphicx} 
\usepackage{preamble}
\usepackage{color}
\usepackage[dvipsnames]{xcolor}
\usepackage{cancel}
\usepackage{orcidlink}

\newcommand{\diff}{\mathrm{d}}

\usepackage{mathtools}

\DeclarePairedDelimiter\floor{\lfloor}{\rfloor}

\usepackage{bm}

\begin{document}

\title{Fractionalization of flux tubes in 3d  and   screening by emergent electric charges in 2d}

\author{Mendel Nguyen \orcidlink{0000-0002-7976-426X}}
\email{mendelnguyen@gmail.com}
\affiliation{Department of Mathematical Sciences, Durham University, Durham DH1 3LP, UK \looseness=-1}
\author{Mithat \"{U}nsal \orcidlink{0000-0002-4875-9978}}
\email{unsal.mithat@gmail.com}
\affiliation{Department of Physics, North Carolina State University, Raleigh, NC 27607, USA \looseness=-1}

\begin{abstract} 
We consider a class of 3d theories with a $ \mathbb Z_n$ magnetic symmetry in which confinement is generated by charge $n$ clusters of monopoles. 
Such theories naturally arise in  quantum antiferromagnets in 2+1, QCD-like theories on $\mathbb R^3 \times S^1$, and $U(1)$ lattice theory with restricted monopole sums.  
A confining string fractionates into $n$ strings which each carry $1/n$ electric flux.  
We construct a twisted compactification  (equivalently periodic compactification with a topological defect insertion) on $\mathbb R^2 \times S^1$ that preserves the vacuum structure. 
Despite the absence of electric degrees of freedom in the microscopic Lagrangian, we show that large Wilson loops are completely/partially screened for even/odd $n$, even when the compactification scale is much larger than the Debye length.  
We show the emergence of fractional electric charges $(\pm 2/n)$ at the junctions of the domain lines and topological defects. 
We end with some remarks on screening vs. confinement. 
 
\end{abstract}

\maketitle
\newpage

\section{I\lowercase{ntroduction}} 
An important aspect of confining gauge theories is the formation of electric flux tubes (stable or metastable) between external probe charges.  
The flux tube emerges as the minimal energy configuration because a uniform spreading of electric flux in the vacuum of a confining theory is energetically much more costly 
\cite{Polyakov:1975rs, Polyakov:1987ez, Wen200710}.

In $d=3$ and $d=2$ dimensions,  
there are two  extremely intriguing phenomena about the nature of flux tubes. 
One is that the structure of the flux tubes is sensitive to the confinement mechanism (eg. magnetic bions vs. monopoles) 
\cite{Polyakov:1975rs, Unsal:2007jx, Unsal:2007vu,Unsal:2008ch}. 
For example, in theories exhibiting the magnetic bion mechanism, such as QCD(adj) on $\mathbb R^3 \times S^1$, the flux tube has a  composite structure \cite{Anber:2015kea}, whereas in the Polyakov model, it does not \cite{Polyakov:1975rs, Polyakov:1987ez}.  
The second phenomenon is that in some 1+1 dimensional gauge theories, external probe charges are fully or partially screened, unexpectedly in the case when the representation of the probe charge is smaller than the representation of the dynamical charges \cite{Gross:1995bp}. 
For example, a probe in the fundamental representation is screened in $d=2$ QCD with adjoint Majorana fermions in the absence of mass or four-fermion deformations \cite{Cherman:2019hbq, Komargodski:2020mxz}. 

In this work, we describe a general class of theories on $\mathbb R^3$ which exhibit flux tube fractionalization into $n$ lines which each carry electric flux $1/n$. 
When compactified on $\mathbb R^2 \times S^1$ preserving adiabatic continuity of the vacuum structure \cite{Unsal:2008ch}, they surprisingly exhibit perfect or partial screening even in the {\it absence} of electric matter fields in the microscopic Lagrangian.   

We consider theories in which confinement is generated by the proliferation of charge $n$ clusters of magnetic monopoles in 3d, with exact (or emergent) magnetic $\Z_n$ symmetry. 
The basic model we have is motivated by examples from condensed matter physics and gauge theory: 
quantum antiferromagnets on bipartite or honeycomb planar lattices \cite{Haldane:1988zz,Read:1989zz,Read:1990zza} 
asymptotically free $SU(2)$ Yang-Mills at $\theta=\pi$, 
gauge theories with matter in the fundamental, adjoint, 3-index symmetric or 4-index symmetric representations formulated on $\mathbb R^3 \times S^1$  \cite{ Unsal:2007jx,Unsal:2008ch, Poppitz:2009tw, Anber:2011de, Sulejmanpasic:2016uwq}, 
quantum dimer models \cite{ Banerjee:2023pnb, Banerjee:2013dda, Banerjee:2015pnt, Banerjee:2014wpa}, 
and $U(1)$ lattice gauge theory with restricted monopole sums  \cite{Sulejmanpasic:2019ytl}.  
In the charge $n$ model, we show that a flux tube fractionates into $n$ lines, and each fractional line carries $1/n$ electric flux, generalizing the effect observed in \cite{Anber:2015kea} for $n=2$.  
 
We then construct a twisted spatial compactification  of the model on $\mathbb R^2\times S^1$  which respects adiabatic continuity \cite{Hayashi:2024yjc, Tanizaki:2022ngt, Unsal:2008ch}  as opposed to thermal (periodic) compactification which has a phase transition 
\cite{Dunne:2000vp, Kovchegov:2002vi, Simic:2010sv}.  
The $S^1$ size, $\ell_2$, is arbitrary. 
However, to show the most dramatic effect, we take $\ell_2$  much larger than the Debye length $\ell_D$ of the magnetic plasma; hence, the Polyakov-like mechanism is {\it always} operative.  
The naive expectation is that confinement should persist.    
We show that despite the absence of any electric degrees of freedom in our microscopic charge $n$ model, the compactified theory, for large Wilson loops, exhibits perfect or partial screening behavior. 
In one interpretation, the theory produces emergent fractional electric charges,  $Q= \pm \frac{2}{n}$. 
As a result, a test charge $+1$ is completely screened by emergent charges for even $n$, and is screened down to $+1/n$ for odd $n$.

\section{B\lowercase{asics of the charge} \boldmath$n$ \lowercase{model}}
The simplest example of a confining string is realized in 3d $U(1)$ lattice gauge theory (Polyakov model) \cite{Polyakov:1975rs}.  
The theory has nonperturbative tunneling events, monopole-instantons, and their proliferation generates a nonperturbative mass 
for the photon. 
In order to describe the effective field theory, one uses abelian duality, where the photon is expressed as a compact scalar (called dual photon), 
$*d \sigma \sim  d a$, and the leading monopole operators are expressed in terms of these as  $  \zeta_1   e^{\pm i \sigma}  \sim  e^{- S_0} e^{\pm i \sigma} $.  
In the Polyakov model, the proliferation of these events generates a Lagrangian   
\begin{align}
   {\cal L}_{\rm 3d}=  \frac{g_3^2}{8 \pi^2} (\del_\mu \sigma)^2  - \zeta_1 \cos \sigma,  \qquad \sigma  \sim \sigma + 2 \pi
  \label{potential1}
\end{align}
An external electric charge $\pm 1$ can be viewed as a vortex of the dual photon, $\oint \vec E . d \vec l = \oint d \sigma= \pm 2 \pi $. 
To find the expectation value of a large Wilson loop $W(C)$ and area law associated with linear confinement, one evaluates the path integral over the dual photon field with property 
 $\oint_{C'} d \sigma=  \pm 2 \pi$ where $C'$ has linking number one with $C$.  
 The profile of the flux tube action density in the transversal direction is the same as that of an instanton in the quantum mechanical problem with dimensionally reduced Lagrangian ${\cal L}_{\rm 1d}= A   {\cal L}_{\rm 3d}$, where $A$ is the area of the loop, and it is just a single lump.

 However, there are many cases  
 in condensed matter systems and lattice or continuum gauge theories in which 
monopole induced terms vanish or do not contribute to the bosonic potential \cite{Read:1989zz,Read:1990zza,Unsal:2007jx,Unsal:2008ch, Banerjee:2023pnb, Banerjee:2013dda, Banerjee:2015pnt}. 
In these situation, confinement and mass gap  are caused by charge $n$ clusters of monopoles such as magnetic bions for $n=2$. 
Motivated by  these  theories, we consider the charge $n$ model: 
  \begin{align} 
  {\cal L}_{\rm 3d}=  \frac{g_3^2}{8 \pi^2} (\del_\mu \sigma)^2  - \zeta_n  \cos n \sigma  + \ldots
  \label{n-model}
\end{align}
where $\zeta_n \sim  e^{-n S_0}$ is proportional to the density of $n$-clusters.  
These systems have either an exact or emergent  $\Z_n$  0-form magnetic symmetry, $\sigma \rightarrow \sigma + \frac{2 \pi}{n}$
descending from the microscopic theory. 
To make the phenomena we reveal  transparent, 
we assume  that \eqref{n-model} has an exact $U(1)$ electric 1-form symmetry, unlike the motivating  theories in which only a subgroup of it is exact, i.e., there is no electric matter hidden in ellipsis.   
We note that \eqref{n-model} is the low-energy description of $U(1)$ lattice gauge theory in which the monopole sum is restricted to  magnetic charges equal to integer multiples of $n$; see Appendix \ref{sec:U1LGT}.

The potential leads to spontaneous breaking of the $\Z_n$ symmetry and there are $n$ degenerate vacua, $|\Psi_j \rangle, j=1, \ldots, n$, where the order parameter acquires a vev 
\begin{align}
\langle e^{i \sigma} \rangle =  e^{ i \frac{2 \pi (j-1)}{n}}, \qquad \sigma_j= \frac{2 \pi }{n}(j-1)
\end{align}
On $\mathbb R^3$, the vacua are gapped, confining, and $n$-fold degenerate.  

\begin{figure}[tbp] 
\begin{center}
\hspace{-0.1cm}
\includegraphics[width=0.4\textwidth]{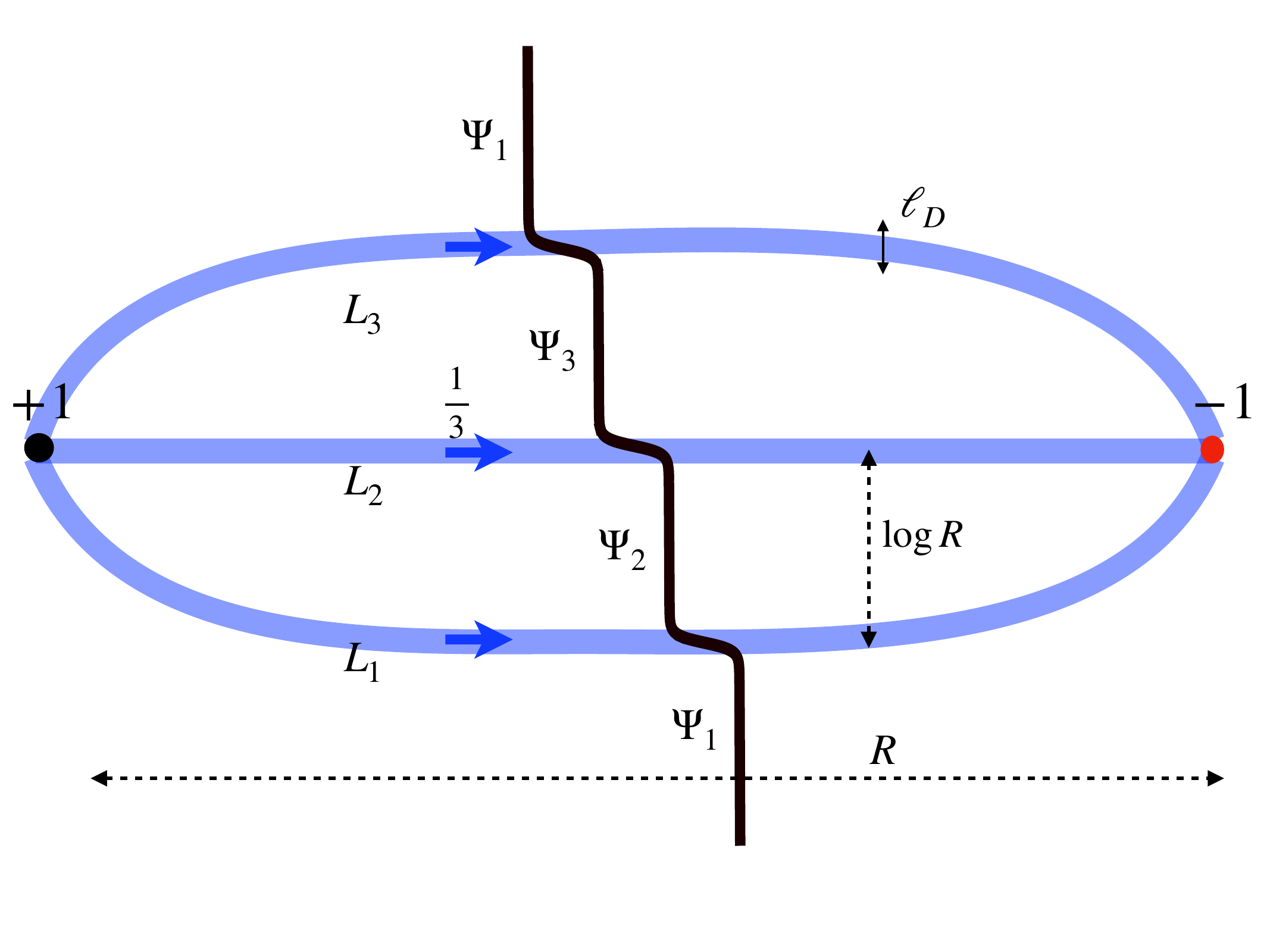} 
\end{center}
\vspace{-1.2cm}
\caption{Electric flux tube in $n=3$ model on $\mathbb R^{3}$ at fixed time slice. It is composed of the domain lines $L_j$,  each of which carry 1/3 flux. The profile of domain lines in the transverse direction is same as the instantons in corresponding quantum mechanics. 
  }
\label{fig:DL}
\end{figure}

\section{D\lowercase{omain lines vs. flux tubes}}


In the 3d long distance theories \eqref{n-model}, the confining strings and the domain lines associated with the breaking of the 0-form $\Z_n$ symmetry are intimately connected. They are both solutions of the same differential equation, but with different boundary conditions.

To determine the domain line profile, let us assume that the line is parallel to the $x_1$ direction (of length $R$, $R \rightarrow \infty$), and in the $x_2$ direction, the field interpolates between two vacua, say, $\sigma_{j}$ and $\sigma_{j+1}$. 
The energy functional  of the domain line  is given by 
\begin{align}
 {\cal E} [\sigma(x_2)] = R  \int dx_2  \biggl[ \frac{g_3^2}{8 \pi^2} \biggl(\frac{ d\sigma}{dx_2}\biggr)^2  + 2 \zeta_n  \sin^2 \frac{n \sigma}{2}  \biggr] 
\end{align}
The domain line $L_j$ is 
associated with the solution of the kink equation:
\begin{align}
\frac{g_3}{2\pi} \frac{d \sigma}{dx_2}= 2 \zeta_n^{1/2} \sin \frac{n\sigma}{2}
\label{kink}
\end{align}
with boundary conditions $\sigma(-\infty)=\sigma_j$, $\sigma(\infty)=\sigma_{j+1}$.  
In the transverse direction to $L_j$, the change in $\sigma$ is smooth and occurs within a Debye length of the system, $\ell_D$. 
The tension of the domain line is: 
\begin{align}
    \Sigma_0= \frac{4 g_3 \zeta_n^{1/2}}{n \pi} 
    \label{tension}
\end{align}

Now,  consider a pair of  external test charges $q_e = \pm 1$  at $x_1=  \mp R/2$, as shown in Fig. \ref{fig:DL}.
Since the theory confines, a flux tube forms between them. 
The profile of the 
flux tube in the $x_2$ direction is found by solving the kink equation \eqref{kink} with   the boundary conditions $\sigma(-\infty)=0$, $\sigma(\infty)= 2\pi$.  
Therefore, as $\sigma$ winds once in the target space, it has to pass through each one of the $n$-vacua located at $\sigma_j$:  
\begin{align}
  \frac{1}{2 \pi}  \oint \diff \sigma =   \frac{1}{2 \pi}  \sum_{j=1}^{n} 
  \int_{ \hat L_j} dx_2  \partial_2 \sigma = 
  \sum_{j=1}^{n} \frac{1}{n} =1
\end{align}
where $\hat L_j $ is the cross-section of the domain line $L_j$.
By abelian duality, $\frac{1}{2 \pi}    \partial_2 \sigma  =E_1$, electric field in the $x_1$ direction.
The monodromy of $\sigma$ implies that the flux tube associated with charge one external probe fractionates into $n$ domain lines, 
\begin{align}
     {\rm Flux \;  tube}   = \prod_{j=1}^{n} L_j 
\end{align}
which each carry fractional electric flux equal to $\frac{1}{n}$ within a domain line of thickness $\ell_D$, 
as shown in Fig.\ref{fig:DL}.  
Finally, the repulsive interaction between consecutive kinks leads 
to a characteristic separation $\log(R)$ between the domain lines in the transverse direction  \cite{Anber:2015kea}. 
The tension of the electric flux tube is the sum of the tensions of the domain lines, $\Sigma_{\rm ft}= n \Sigma_0$.

\section{ A\lowercase{rea vs. perimeter law on}  \boldmath$\mathbb R^2 \times S^1$}
The theory on $\mathbb R^3$ confines for any finite  value of $n$, as in the $n=1$ case, the Polyakov model.   
By construction, the theory does not have any dynamical electric charges.  The thermal compactification of the theory undergoes a 
(reversed) BKT  phase transition at a critical temperature $T_c =2 g_3^2 n^2/\pi$, above which the monopole operators become irrelevant, and the dual photon becomes massless \cite{Dunne:2000vp, Kovchegov:2002vi, Simic:2010sv}.
 
To preserve adiabatic continuity of the vacuum, we consider a spatial compactification using a charge-conjugation twisted boundary condition, 
 \begin{align}
    \sigma(x_1, x_2+ \ell_2, x_3)=-\sigma(x_1, x_2, x_3) 
    \label{bc}
 \end{align}
following similar ideas as in gauge theory on 
$\mathbb R^3 \times S^1$ \cite{Unsal:2007jx,Unsal:2007vu, Unsal:2008ch}.
For $n$ odd, the only vacuum state that can fill up the whole space is $\Psi_1$, while for $n$ even, either $\Psi_1$ or $\Psi_{n/2+1}$ are possible. 
We will observe that there are some remarkable emergent phenomena taking place in this setup. 
We describe these in three complementary perspectives:  a) using the twisted boundary condition, 
b) using the periodic boundary condition with a topological defect insertion, c) using the transmutation of monopole fields  
into vortex fields.

\begin{figure}[tbp] 
\begin{center}
\hspace{-0.1cm}
\includegraphics[width=0.5\textwidth]{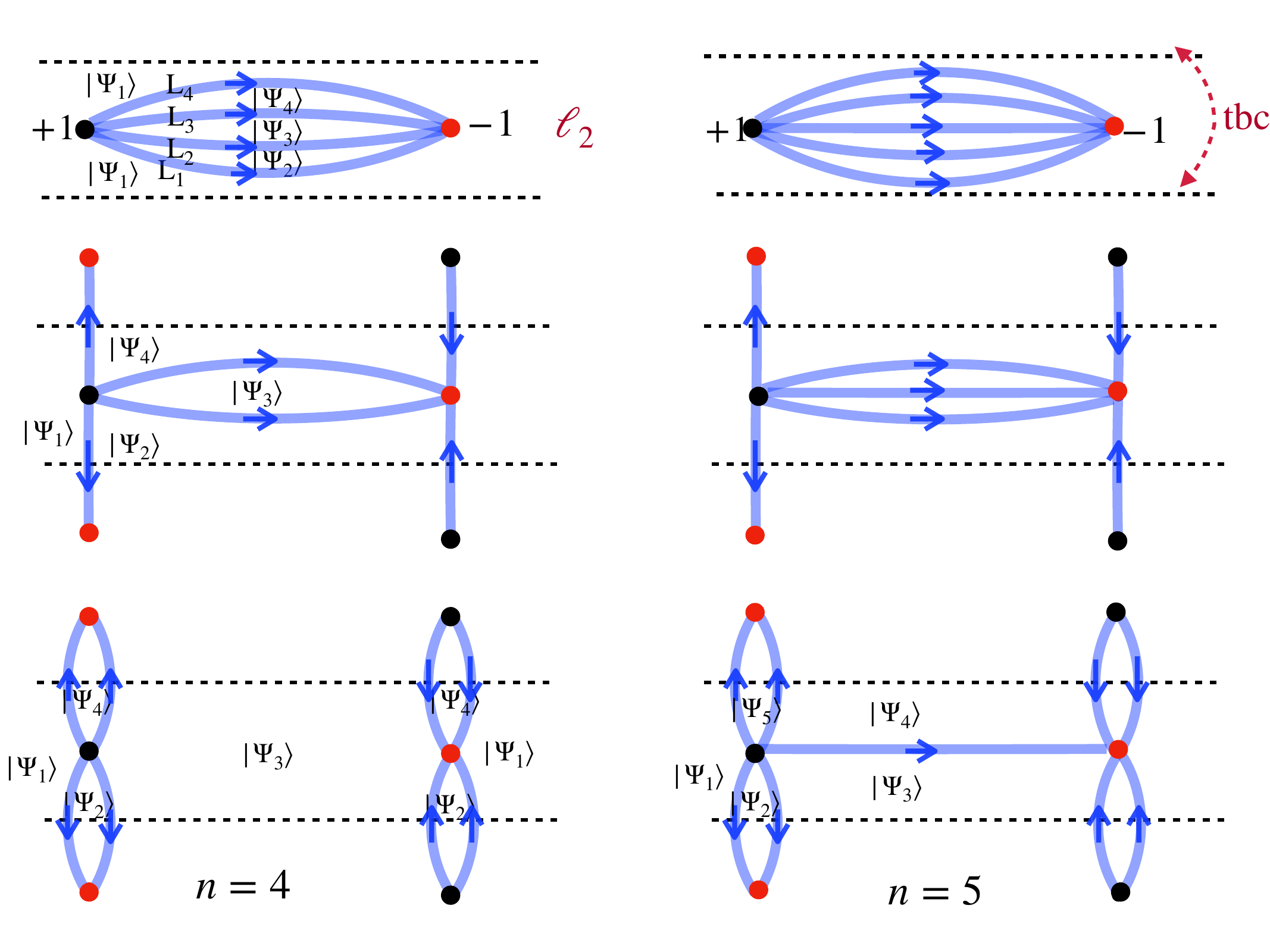} 
\end{center}
\vspace{-0.5cm}
\caption{The structure of  electric flux tube in $n=4,5$ models on   $\mathbb R^2 \times S^1$ with $\ell_2 \gg \ell_D$.   The magnetic charges proliferate in the vacuum. However,  
for even $n$,  the area law  turns to perimeter law at large distance.  For odd 
$n$, area law survives, but tension is reduced by a factor of $n$. 
  }
\label{fig:frac}
\end{figure}

Let $[0, \ell_2)$ denote the fundamental domain $D_0$.  
The sign of the  field is flipped in adjacent domains \cite{Hayashi:2024yjc}.  
As a result, an electric charge configuration with $ \frac{1}{2 \pi} \oint d\sigma = +1$ in  $D_0$ maps to    
$ \frac{1}{2 \pi} \oint d\sigma = -1$ in $D_{\pm1}$, and alternate from there on. 
A pair of $\pm$ electric charges at $p_1, p_2$ is mapped to a pair with $\mp$ charges in adjacent slabs, located at  $p_1 \pm \ell_2 \hat{e}_2, p_2 \pm \ell_2 \hat{e}_2$.   
Thus, the composite electric flux tube emanating from $p_1$ and terminating at $p_2$ on $\mathbb R^3$ can have different prospects on $\mathbb R^2 \times S^1$.

There are  $ \floor*{\frac{n}{2}} +1$  topologically distinct configurations that satisfies monodromy conditions around $p_1$ and $p_2$, see Fig. \ref{fig:frac}. 
Of the $n$ domain lines (fractional  $e$-flux tubes), $n-2 k$ of them, where $k=0, \ldots, \floor{n/2}$,
go from $p_1$ and to $p_2$, $k$ of them $(L_1,\ldots L_k)$ go from $p_1$ 
to $p_1 - \ell_2 \hat e_2$, and the remaining $k$ $(L_n,\ldots L_{n-k+1})$ go from 
$p_1$ to $p_1 + \ell_2 \hat e_2$.  
These configurations are consistent with the twisted boundary condition.  

For even $n$, and  $k=n/2$, there is no line going from $p_1$ to $p_2$, and in this configuration, it does not cost any energy to separate charges at $p_1$ and $p_2$ further apart.  
For odd $n$, it is impossible to get rid of the line $L_{(n+1)/2}$ going from $p_1$ to $p_2$, and linear confinement persists.
The bottom row of Fig. \ref{fig:frac} displays the minimal energy configurations for $n=4,5$.

\begin{figure}[tbp] 
\begin{center}
\includegraphics[width=0.5\textwidth]{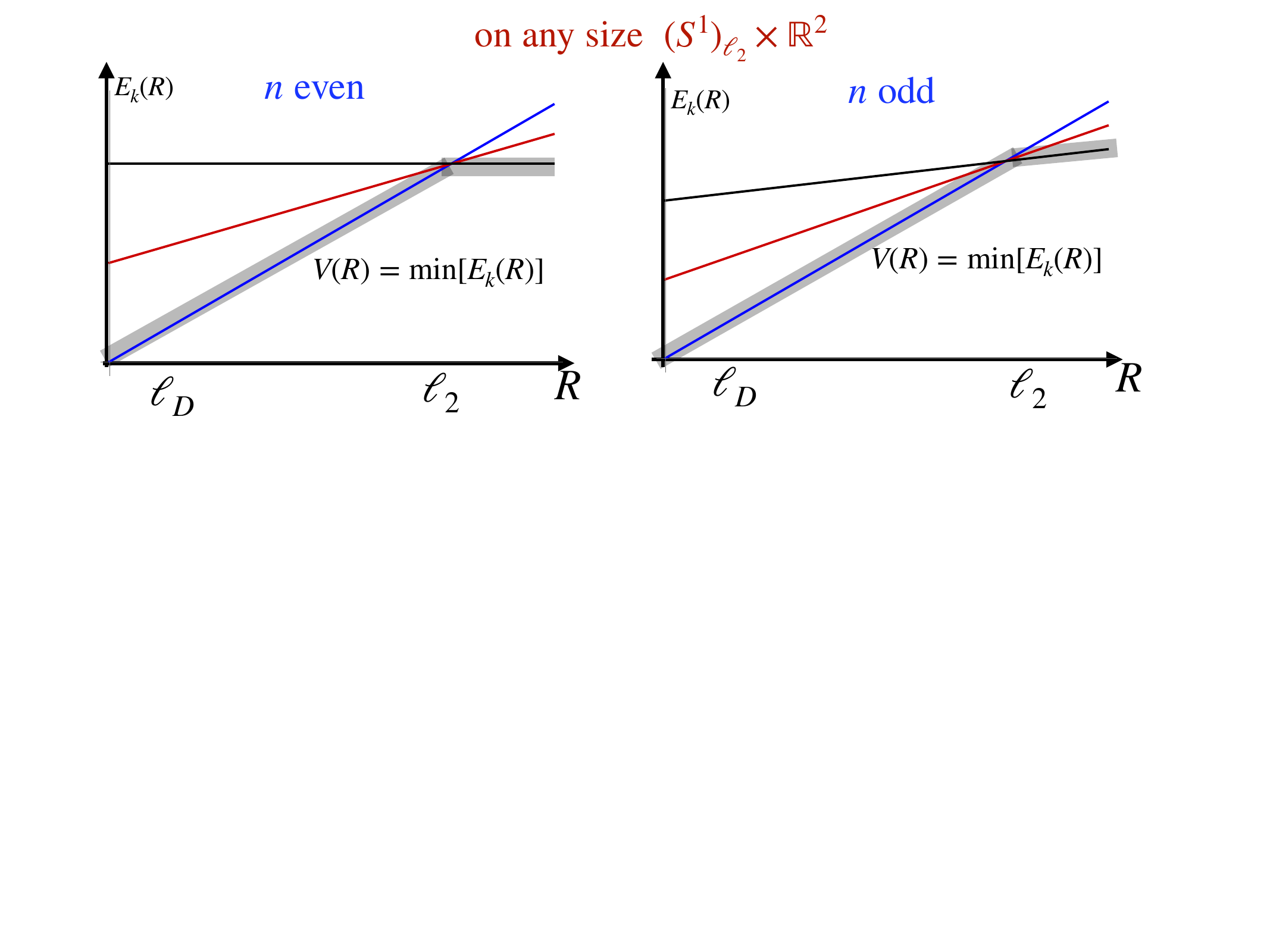} 
\end{center}
\vspace{-4cm}
\caption{Potential between two  test charges for twisted compactification on  $\mathbb R^2 \times S^1$ is shown on gray band for $n=4, 5$.   
The scale at which the slope changes tends to infinity as $\ell_2 \rightarrow \infty$. 
  }
\label{fig:tension}
\end{figure}

The energies of these configurations are given by:
\begin{align}
    E_k= \Sigma_0 [ (n-2k) R + 2 k \ell_2 ], \quad k= 0, 1, \ldots, \floor*{\frac{n}{2}} 
\end{align}
The expectation value of a rectangular Wilson loop in the $(x_1, x_3)$ plane with extension $T$ in $x_3$ much larger than $R$ in $x_1$   
[area  $ A(C) =RT$, perimeter $P(C) \approx 2T$]  
is  given by a sum over these configurations: 
\begin{align}
\langle W(C) \rangle  &= \sum_{k=0}^{\floor*{{n}/{2}}} e^{ -E_k T }  \cr 
&= \sum_{k=0}^{\floor*{{n}/{2}}}    e^{ -\Sigma_0 (n-2k) A(C)  - (k \Sigma_0 \ell_2) P(C)} 
\label{WL}
\end{align}
The potential energy between the two test charges is
\begin{align}
V(R) = \lim_{T \rightarrow \infty} -\frac{1}{T}  \log \langle W(C) \rangle
   = R \min_k [E_k]  
\end{align}
as shown in Fig.\ref{fig:tension}, 
leading to string tensions 
\begin{align}
\Sigma_{\rm ft}
   = \left\{ \begin{array}{lll}
        n \Sigma_0,  & R < \ell_2, & \forall n  \\
         \Sigma_0, & R > \ell_2, & {\rm odd} \;  n  \\ 
         0,  & R > \ell_2, &  {\rm even} \;  n
   \end{array} 
   \right.
\label{screen}
\end{align}
What does this imply?  
For $R   \lesssim  \ell_2$, both for odd and even $n$, we observe linear confinement, with string tension $n \Sigma_0$. 
However, for $R  \gtrsim  \ell_2$, the theory with even $n$ exhibits complete screening, while for odd $n$, the theory exhibits linear confinement albeit with a reduced string tension $\Sigma_0$. 
One can call this imperfect screening. 

\section{f\lowercase{ractional screening}}

The characteristic feature of \eqref{WL} is the superposition of area and perimeter law terms, which can be interpreted as  follows. 
The  $ e^{ -\Sigma_0 (n-2k) A(C)  - (k \Sigma_0 \ell_2) P(C)}$ term ($k=  1, \ldots, \floor*{\frac{n}{2}}$)  in  \eqref{WL} would arise if one had fractionally charged massive fields with 
\begin{align}
 m^*_k = k \Sigma_0 \ell_2,  \qquad  Q_k =- \frac{2k}{n}
\end{align}
where the mass is $k$-many domain line tensions multiplied with the size $\ell_2$. 
Therefore,  charge one electric probes can be totally screened for even $n$. For odd $n$,  all but a $1/n$ fraction of it can be screened. 
These perfect and imperfect screenings occur in the absence of electric degrees of freedom in the microscopic Lagrangian. 
We would like to explain the emergence of these fractional massive charges.

\begin{figure}[tbp] 
\begin{center}
\hspace{-0.1cm}
\includegraphics[width=0.5\textwidth]{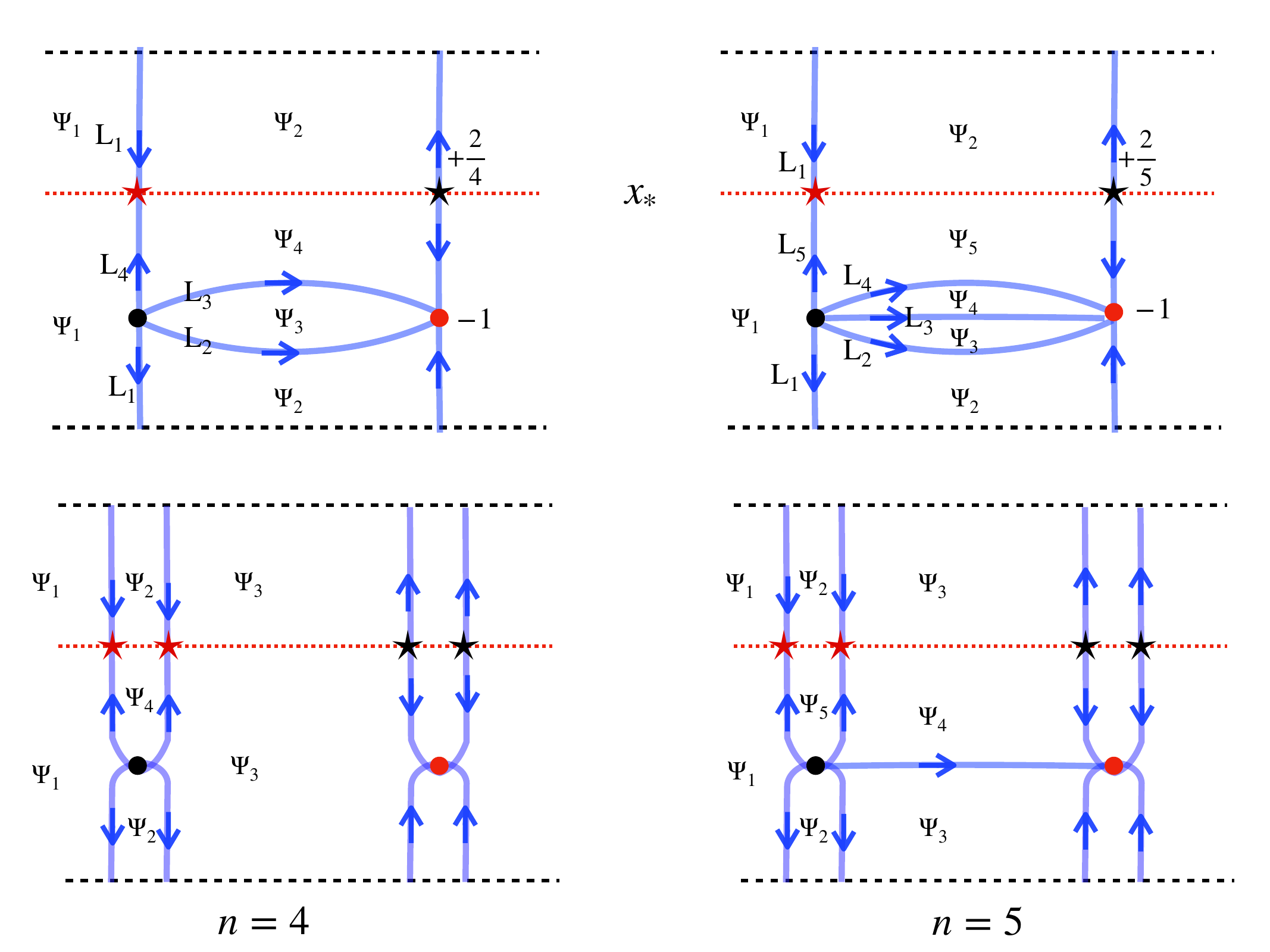} 
\end{center}
\vspace{-0.5cm}
\caption{Converting twisted boundary conditions to periodic ones leads to 
a defect line at some arbitrary point $x_* \in S^1$.  
 The point where the lines $L_j$ fuses with $L_{n+1-j}$ 
carries $-2/n$ electric charge that soaks up the electric flux of the lines.  
  }
\label{fig:defectfrac}
\end{figure}

On $\mathbb R^2 \times S^1$, turning twisted  to periodic boundary conditions by a field redefinition requires insertion of topological defect lines (or surfaces in spacetime) at  position $x_{*} \in S^1$ (red-dotted lines in Fig.\ref{fig:defectfrac}.) 
Traversing the defect line, one transforms the states and domain lines  by $\mathbb Z_2$ conjugation: 
\begin{align}
   \Psi_j \rightarrow \Psi_{n+2-j}, \qquad L_j \rightarrow L_{n+1-j} 
\end{align}
The junction of the defect line with the domain lines $(L_j, L_{n+1-j})$ must absorb 
the electric flux of the domain lines, and therefore carry a charge $-2/n$. 
\begin{align}
    Q_{\rm d}(L_j, L_{n+1-j}) = - \frac{2}{n}
\end{align}
As shown in Fig.\ref{fig:defectfrac}, for even $n$, the  fusion of  $L_1, \ldots, L_{ \lfloor \frac{n}{2} \rfloor}$ with  $L_n, \ldots, L_{ n+1 - \floor{\frac{n}{2}}}$ carries $-1$ and screens the external charge completely. 
For odd $n$, the same combination carries electric charge $-1+ \frac{1}{n}$ and screens all but $1/n$ of external charge.  
Hence, confinement survives for odd $n$.

  There is something surprising even in the Polyakov model, $n=1$ \cite{Polyakov:1975rs}. 
  Consider charge-$q$  probes. 
On $\mathbb R^3$, flux tube is composite of $q$ charge-1 flux tubes!   
On  large $\mathbb R^2 \times S^1$ with 
  \eqref{bc},    Wilson loops with  $R < \ell_2$ exhibit string tension  with $q \Sigma_0$.  For  
  $R > \ell_2$,  $q$-even Wilson loops exhibit perimeter law  , and all $q$-odd loops exhibit 
  area law with  tension  $\Sigma_0$, as in center-vortex scenario \cite{Greensite:2003bk}.  The emergent charges in Fig.\ref{fig:defectfrac} are $\mp 2$
  providing a connection to center-vortex mechanism  \cite{Hayashi:2024yjc, Hayashi:2024psa}.


\section{C\lowercase{harge} \boldmath$n$ \lowercase{monopole to  local vortices}}  
Now we explain the emergence of the perimeter-law for even $n$, and its non-emergence for odd $n$ under the twisted compactification from the perspective of the monopoles, similar to the constructions in \cite{Hayashi:2024yjc, Hayashi:2024psa, Guvendik:2024umd}.   
Because of the twisted boundary condition, the monopole charges alternate between the adjacent domains and  the field of a monopole  turns  into the field of an infinite array of alternating charges.  
This is mathematically identical to the electric  field of a charged  particle between two parallel conducting plates \cite{Schwinger}.  
The  potential due to an alternating array of charge $\pm n$ monopoles is
\begin{align}
V(\rho, x_2) =  \frac{n}{4\pi} \sum_{k \in \Z}   
       \frac{1}{ r_k^{+} } 
            -  \frac{n}{4\pi} \sum_{k \in  \Z}      \frac{1}{ r_k^{-} } \; . 
\end{align}  
where $r_k^{\pm}  = \sqrt {(x_2 - \ell_2 (2k + \alpha) )^2 + \rho^2} $ where $\alpha=0, 1/2$ for $+, -$ charges respectively. 
One can write the potential in a form that makes its long-distance properties more manifest:  
\begin{align}
 V(\rho, x_2)  =   \frac{1}{2 \pi  \ell_2}     
   \sum_{m=1,3,..}     
       K_0 \Big(\frac{ \pi}{\ell_2} m \rho  \Big)
      \cos \Big(\frac{ \pi}{ \ell_2} m x_2 \Big), 
\end{align} 
The  algebraic long range Coulomb potentials $( \pm 1/r)$ sum up to an exponentially decaying field at distances $\rho >  \ell_2 /\pi$.  
This implies that the magnetic flux of a monopole collimate into upward and downward directed magnetic flux tubes of thickness $\ell_2/ \pi$.
 
Let us determine the value of a classical Wilson loop in the $(x_1, x_3)$ plane.  
A monopole  of charge $n$   at  a  point $x$  yields 
 \begin{align}
W_{\rm cl.} (C) = e^{i \oint_C A_m} =  \left\{
  \begin{array}{ll} 
e^{i n \times  \frac{1}{2} \Omega_D(x)},  & \ell_2  \gg  R    \\  
e^{i   n \times \pi  \Theta_D(x)}, &  \ell_2 \ll R 
\end{array}  \right.
\end{align} 
For $\ell_2  \gg  R$, the result is the same as in $\mathbb R^3$. 
The Wilson loop measures $n \times  \Omega_D(x)/2$,  $n$ times half of the solid angle    subtended by the 
oriented  minimal surface $D$  relative to the  position of the monopole $(\partial D=C)$.  
For  $\ell_2  \ll  R$, the 2d observer sees the effect of the monopole as a vortex. The flux of the vortex is given by $\pi n \Theta_D(x)$ 
where  $\Theta_D(x)$ is the step function equal to  one for  $(x_1, x_3) \in C$ and zero otherwise.

%

For odd $n$, the vortex  flux is nontrivial, an odd multiple of $\pi$; hence, proliferation of vortices in 2d generates area law of confinement as shown in \cite{Tanizaki:2022ngt}. 
However, for even $n$, the vortex flux is an even multiple of $\pi$, and has no effect on the Wilson loop. 
Hence, for a macroscopic 2d observer, the large Wilson loops exhibit  perimeter  law for even $n$ consistent with the large $R$ limit  of  \eqref{screen}.

 \section{C\lowercase{onclusion}} 
It is for certain that the charge $n$ model (like the Polyakov model) exhibits nonperturbative mass generation and confinement on $\mathbb R^3$. 
Let us say we compactified the theory on  $\mathbb R^2 \times S^1$ with $\ell_2 \gg \ell_D$. 
The mass gap, vacuum degeneracy, etc. are still dictated by proliferation of charge $n$ monopoles as on $\mathbb R^3$.  
Wilson loops with  $R < \ell_2$ still obey area law and exhibit the existence of flux tubes. 
However, the standard definition of confinement asserts that  we need to look to asymptotically large loops and they must exhibit area law, rather than perimeter law.  
In our even $n$ model, a perimeter law kicks in for $R \geq \ell_2$ where  $\ell_2 \gg \ell_D$ arbitrarily large. 
What should we call this phase? 
Confining or screening? 
 
Unlike real QCD, in this case, there is an exact electric 1-form symmetry, and we should be able to give an unambiguous answer.  
The standard definition tells us to call it screening if the 1-form symmetry is broken. 
On the other hand, there exists a flux tube and it persists for any $R \leq \ell_2$, and recall that  in our analysis, we are allowed to take $ \ell_2 $ arbitrarily large. 
Any local observer (by seeing the flux tube)  will say that the theory is confining.



\acknowledgments
We thank Mohamed Anber, Michelle Caselle, Aleksey Cherman, David Gross, Canberk G\"uvendik, Yui Hayashi, Zohar Komargodski, Sahand Seifnashri, and Yuya Tanizaki for various 
conversation about this work. 
M.\"U. is supported by U.S. Department of Energy, Office of Science, Office of Nuclear Physics under Award Number DE-FG02-03ER41260.

\appendix

\section{Theories that lead to charge \boldmath$n$ monopole potential at leading order} 

Below, we describe physical situations in which the potential in the dual photon Lagrangian  \eqref{n-model} arise as the leading bosonic potential.  
There are two physical mechanisms by which this occurs. 
The mechanism in $SU(2)$ deformed Yang-Mills on $\mathbb R^3 \times S^1$, quantum anti-ferromagnets, and $U(1)$ lattice gauge theory with restricted monopole sums rely on destructive interference between monopoles. 
The mechanism that takes place in $SU(2)$ gauge theory with fermionic matter is sourced by fermionic zero mode induced clustering.  
Let us review both at the conceptual level.

In $SU(2)$ deformed Yang-Mills on $\mathbb R^3 \times S^1$, once $SU(2)$ Higgses to $U(1)$, there are two types of monopole operators, $M_1 \sim e^{i \sigma + i \theta/2} $ and $M_2 \sim e^{-i \sigma + i \theta/2}$, with magnetic and topological charge $(+1, 1/2)$ and $(-1, 1/2)$. 
Proliferation of monopoles and antimonopoles induces a potential $\cos( \frac{\theta}{2}) \cos(\sigma) $ which dies off at $\theta=\pi$. 
Essentially, $\bar M_2$ cancels the effect of $M_1$, $(1+ e^{i \pi})e^{i \sigma}=0$.  
At second order in the cluster expansion, there are magnetic bions $[M_1 \bar M_2] \sim e^{2 i \sigma}$ with no theta dependence. 
Proliferation of bions generates the potential  \eqref{n-model}with $n=2$. 

Two dimensional quantum antiferromagnets on bipartite square lattices also exhibit similar behaviour. 
The properties of the Coulomb plasma vary periodically with the spin of states on each site depending on  $2S = 0,1,2,3 \; (\rm mod 4)$.  
The monopoles tied with certain sublattices acquire different Berry phases, such that the proliferation of monopoles with charge $m$ yields:
 \begin{align}
  e^{-m S_0} e^{i m \sigma} \sum_{\zeta_s=0, 1,2,3} e^{i \pi S  m \zeta_s} 
 \end{align}
Therefore, for $2S = 0,1,2,3 \; (\rm mod 4)$, the leading operator that will be generated in the effective Lagrangian  will have $n= 1,4,2,4$, which is the ground state degeneracy for the system. 
Similarly, on a honeycomb lattice, the Berry phases are $e^{i \frac{\pi 4 S}{3}  m \zeta_s}$ and the leading contribution to potential for $S=1/2$ is  $n=3$. 
Hence, the ground state is three-fold degenerate. 

In $SU(2)$  gauge theory with fermionic matter, the crucial element that ultimately leads to the same effect is fermion zero modes. 
We consider the theories 
with one Weyl fermion, in  adjoint, 3-, or 4-index symmetric representations.   
The 3-index  theory is chiral.  
All three theories have at the classical level a $U(1)$  axial symmetry. 
Since the 4d instanton has, respectively, $I_{4d}= 4, 10, 20 $ fermi zero modes, the discrete chiral symmetry of the quantum theory is $\Z_4, \Z_{10}, \Z_{20} $ respectively.  
We expect these symmetries to be broken down to $\Z_2$, hence a ground state degeneracy of $d= 2, 5, 10$ respectively on $\mathbb R^3 \times S^1$.   
In each case, the number of zero modes distributes to $M_1$ and $M_2$ monopole-instantons. 
We have, for adjoint $(I_1, I_2)=(2,2)$,  for 3-index, 
$(4,6)$, for 4-index,  $(6, 14)$  fermi zero modes. 
The monopole operators are of the form: 
\begin{align}
     M_1 \sim  e^{-S_0} e^{i \sigma } \psi^{I_1},  \quad  M_2 \sim e^{-S_0} e^{-i \sigma } \psi^{I_2}  
\end{align}
Hence, monopoles cannot generate a purely bosonic potential. 
To generate something purely bosonic, the fermi zero modes  must be soaked up between the constituents of the  cluster. 
The minimal objects that satisfy this are 
\begin{align}
   [M_1^{I_2/2}  \overline M_2^{I_1/2}] \sim    e^{- \frac{(I_1 + I_2)}{2} S_0  } e^{i  \frac{(I_1 + I_2)}{2}  \sigma } 
\end{align} 
For adjoint, there are magnetic bions $[M_1 \bar M_2] \sim e^{2 i \sigma}$.  
Similarly, for 3- and 4-index symmetric representations,  the leading  objects that contribute to the bosonic potential in the cluster  expansion  are $[M_1^3 \bar M_2^2]$,  and $[M_1^7 \bar M_2^3]$, and they generate the potentials \eqref{n-model} with $n=2, 5, 10$ respectively.

\section{3d U(1) lattice gauge theory with charge \boldmath$n$  monopoles or monopole-clusters \label{sec:U1LGT}} 

We describe, for any given $n$, a general construction in 3d $U(1)$ lattice gauge theory that produces the leading effective potential $\cos(n\sigma)$ induced by $n$-clusters of charge $1$ monopoles.  

\newcommand{\wilson}{S_{\text{\sc w}}}
\newcommand{\villain}{S_{\text{\sc v}}}
Let us start by recalling that in the weak coupling regime of the 3d $U(1)$ lattice gauge theory as formulated by Wilson,
\begin{equation}
    S = \frac{1}{2e^2} \sum_{x,\mu\nu} (1- \cos F_{x,\mu\nu})
\end{equation}
where $F_{x,\mu\nu} \equiv \Delta_\mu A_{x,\nu} - \Delta_\nu A_{x,\mu}$ is the field-strength, we may pass over to the so-called Villain theory
\begin{equation}
    S = \frac{1}{4e^2} \sum_{x,\mu\nu} (F_{x,\mu\nu} + 2\pi m_{x,\mu\nu})^2
    \label{eq:villain}
\end{equation}
in which we further require an integer-valued dynamical variable $m_{x,\mu\nu}$ associated with plaquettes, reflecting the periodicity of the Wilson action.
Physically, $2\pi m_{x,\mu\nu}$ is interpreted as the magnetic flux through the plaquette $(x,\mu\nu)$, and the magnetic charge in a region is found by summing $m_{x,\mu\nu}$ over the boundary.
Thus, the magnetic charge density $q_z$ (in units of $2\pi$) is 
\begin{equation}
    q_z = \tfrac12 \epsilon_{\lambda\mu\nu} \Delta_\lambda m_{x,\mu\nu}
\end{equation}
As shown by Polyakov \cite{Polyakov:1975rs}, the theory may be reformulated in such a way that the vector potential is eliminated and the magnetic charges serve as the basic dynamical variables:
\begin{equation}
    S = \frac{(2\pi)^2}{2e^2} \sum_{\smash{z,z'}} q_z G_{z,z'} q_{z'}
\end{equation}
where $G_{z,z'}$ is the Green's function of the discrete Laplacian.
For a configuration where only finitely many sites $z_r$ have $q_z$ nonzero, and where each separation $|z_r - z_s|$ with $r\neq s$ is large, we can approximate the action by
\begin{equation}
    S = \frac{(2\pi)^2}{2e^2} \sum_{r<s} q_r G(z_r - z_s) q_s + \sum_r S_0 (q_r)^2
\end{equation}
where $S_0 \equiv (2\pi)^2 G_0/2e^2$, $q_r \equiv q_{z_r}$ and we have replaced the lattice Green's function by the continuum Green's function $G(z)$.
This is recognized as the classical magnetostatic energy of a gas of various species of monopoles, with the chemical potential of each species proportional to (minus) the square of the charge. 
It is evident that the dominant contribution to the path integral -- or grand canonical partition function in the monopole gas language -- comes from the species with charge $\pm 1$.
Summing only over these species, one arrives at an effective scalar field action of the form
\begin{equation}
    S = \int d^3 z \biggl[ \frac{e^2}{2(2\pi)^2}(\partial_\mu \sigma)^2 - 2 e^{-S_0} \cos \sigma \biggr]
\end{equation}

\begin{figure}[tbp] 
\begin{center}
\includegraphics[width=0.5\textwidth]{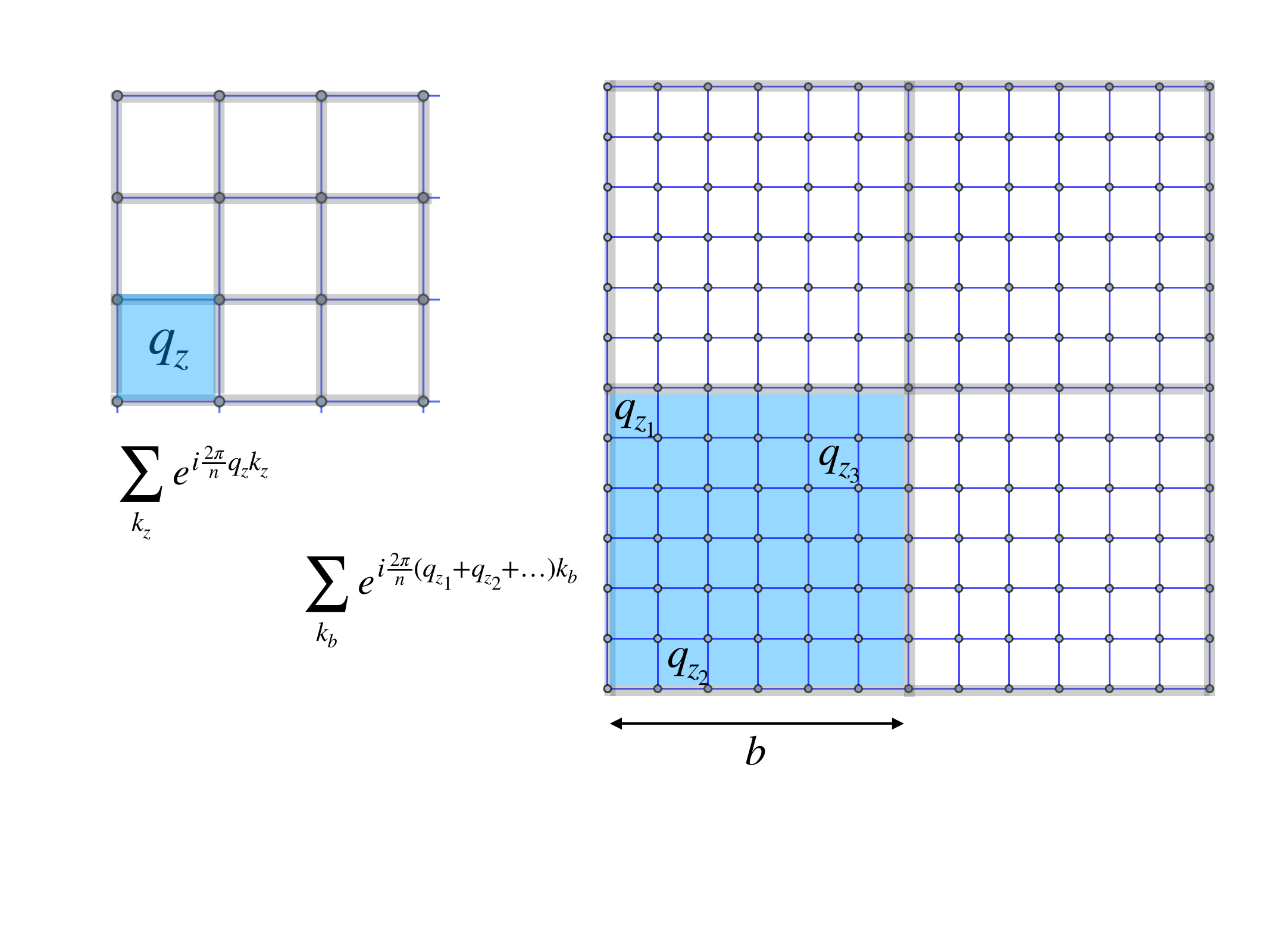} 
\end{center}
\vspace{-2.0cm}
\caption{2d slice of 3d lattice. Imposing the charge-$n$ constraint via coarser blocks, the action of the leading configuration changes from $S =S_0 n^2$ for $b=1$  to $S =S_0 n$ for $b$ larger than magnetic bion size.   
  }
\label{fig:constraint}
\end{figure}

To construct a model in which the leading monopole-induced potential is $\cos(n\sigma)$ with $n>1$, we employ the following device which produces what is known as a modified Villain formulation \cite{Sulejmanpasic:2019ytl, Gorantla:2021svj}.
We introduce an integer-valued field $k_z$ in addition to the vector potential and flux fields of the standard Villain theory and take the action
\begin{equation}
    S = \frac{1}{4e^2} \sum_{x,\mu\nu} (F_{x,\mu\nu} + 2\pi m_{x,\mu\nu})^2 - \frac{2\pi i}{n} \sum_z k_z q_z
    \label{eq:mvillain}
\end{equation}
On summing out $k_z$, the last term drops out and the remaining sum over $m_{x,\mu\nu}$ is constrained by the condition that $q_z$ should always be a multiple of $n$.
Thus, in the monopole gas description of the system, only species with charge a multiple of $n$ contribute, the contribution from other charges canceling out due to the destructive interference in the sum over $k_z$. 
By literally repeating the preceding arguments, one arrives at the effective scalar field action
\begin{equation}
    S = \int d^3 z \biggl[ \frac{e^2}{2(2\pi)^2}(\partial_\mu \sigma)^2 - 2 e^{-S_0 n^2} \cos (n\sigma) \biggr]
\end{equation}
However, a deficiency of this construction, insofar as it differs from the models previously described, is that the proliferating charge $n$ objects here cannot be interpreted as $n$-clusters of charge $1$ monopoles.
In particular, the charge $n$ fugacity $\smash{e^{-S_0 n^2}}$ is not the $n$th power of the charge $1$ fugacity $e^{-S_0}$ but much smaller: $ \smash{e^{-S_0 n^2}} \ll e^{-S_0 n}$.

Nevertheless, we can adapt the modified Villain approach so that the effective potential $\cos(n\sigma)$ will indeed be produced by $n$-clusters of charge $1$ monopoles.
We can take the same action \eqref{eq:mvillain}, but now we make a blocking of the lattice and require that for each block $b$, all $k_z$s with $z \in b$ take the same value $k_b$.
Then summing out $k_b$ restricts the sum over $m_{x,\mu\nu}$ to configurations with net charge a multiple of $n$ within each block, while still allowing configurations with net charge $<n$ within an elementary cube.
If the block size is $b^3 \sim n$, we can accommodate $n$ charge-1 monopoles within the block, but one can show that this configuration is still order $n^2$ action because of the mutual interactions of order $\frac{1}{e^2} \frac{n(n-1)}{r_{ij}}$. 
However, if the block size is larger than the magnetic bion size $r_b \sim r_m/(e^2 a)$ \cite{Unsal:2012zj}, then the action of the $n$-cluster is the sum of the actions of 1-monopole events at leading order in semiclassics,  $S_n = S_0 n $.  Now, $\frac{1}{e^2} \frac{n(n-1)}{r_{ij}} \sim n(n-1)  $ is independent of 
 $e^2$ and does not contribute to the classical action at leading  $1/(e^2 a)$ order.

Let us now be a bit more explicit about how the effective field theories are derived from the modified Villain models.
The first step is to decompose the magnetic flux variables $m_{x,\mu\nu}$ as follows:
\begin{equation}
    m_{x,\mu\nu} = \Delta_\mu \alpha_{x,\nu} - \Delta_\nu \alpha_{x,\mu} + \epsilon_{\lambda\mu\nu} \Delta_\lambda \phi_z
    \label{eq:helmholtz}
\end{equation}
where $\phi_z$ is determined by the discrete Poisson equation
\begin{equation}
    \Delta_\lambda \Delta_\lambda \phi_z = q_z 
\end{equation}
(This is just the discrete analog of the Helmholtz decomposition of 3d vector fields in continuum.)
Substituting the decomposition \eqref{eq:helmholtz} for $m_{x,\mu\nu}$ in \eqref{eq:villain}, and performing the field redefinition $A_{x,\mu} \to A_{x,\mu} - \alpha_{x,\mu}$, we find
\begin{multline}
    \frac{1}{4e^2} \sum_{x,\mu\nu} (F_{x,\mu\nu} + 2\pi \epsilon_{\lambda\mu\nu} \Delta_\lambda \phi_z)^2
    \\
    = \frac{1}{4e^2} \sum_{x,\mu\nu} (F_{x,\mu\nu})^2 + \frac{(2\pi)^2}{2e^2} \sum_z (\Delta_\lambda \phi_z)^2
    \\
    = \frac{1}{4e^2} \sum_{x,\mu\nu} (F_{x,\mu\nu})^2 + \frac{(2\pi)^2}{2e^2} \sum_{\smash{z,z'}} q_z G_{z,z'} q_{z'}
\end{multline}
We see that the vector field decouples from the magnetic charges, and we neglect it henceforth.
Furthermore, the sum over the fluxes $m_{x,\mu\nu}$ effectively becomes a sum over the charges $q_z$. 
The action together with the Lagrange multiplier term now reads
\begin{equation}
    S = \frac{(2\pi)^2}{2e^2} \sum_{\smash{z,z'}} q_z G_{z,z'} q_{z'} - \frac{2\pi i}{n} \sum_z k_z q_{z}
\end{equation}

To pass to the scalar field description, we perform a cluster expansion of the partition function.
Let us separate out the $z=z'$ part of $G_{z,z'}$ and write
\begin{equation}
    S = \frac{(2\pi)^2}{2e^2} \sum_{\smash{z,z'}} q_z \widehat G_{z,z'} q_{z'} + S_0 \sum_z (q_z)^2 - \frac{2\pi i}{n} \sum_z k_z q_z
\end{equation}
where we have put $\widehat G_{z,z'} \equiv G_{z,z'} - G_0 \delta_{z,z'}$.
The partition function can then be expressed as the Gaussian integral
\begin{multline}
    Z = \sum_{\smash{\{k_b\}}}\int [d\sigma_z] \exp \biggl[ - \frac{e^2}{2(2\pi)^2} \sum_{\smash{z,z'}} \sigma_z (\widehat G^{-1})_{z,z'} \sigma_{z'} \biggr] \\
    \times \prod_z \sum_{q_z} \exp\biggl[-S_0(q_z)^2 + i q_z \sigma_z + \frac{2\pi i}{n} k_z q_z \biggr]
    \label{eq:mvillainpartition}
\end{multline}
[Note that $(\widehat G^{-1})_{z,z'}$ is a short-range translation-invariant function that, up to higher derivatives, goes to the Laplacian in the continuum.]
Putting $\zeta \equiv e^{-S_0}$ and $\omega \equiv e^{2\pi i /n}$, we rewrite the second line in \eqref{eq:mvillainpartition} as
\begin{equation}
    \prod_{\smash{b}} \prod_{\smash{z\in b}}(1 + \sum_{\smash{q=1}}^\infty \zeta^{q^2} \omega^{qk_b} e^{i q \sigma_z} + \sum_{\smash{q = 1}}^\infty \zeta^{q^2} \omega^{-qk_b} e^{-i q \sigma_z})
\end{equation}
In the product over $z \in b$, the first terms of nonzero charge in the $\zeta$ expansion that survive the summation over $k_b$ are the charge $\pm n$ terms
\begin{equation}
    e^{\pm i \sigma_{z_1}} \ldots e^{ \pm i \sigma_{z_n}} \sim e^{\pm i n \sigma_b}
\end{equation}
($z_1,\ldots,z_n \in b,\ z_i \neq z_j$) at order $\zeta^n$.
These operators then provide the leading contribution $\propto e^{-nS_0} \cos (n\sigma)$ to the effective potential.

From a symmetry perspective, the absence of $\cos(q\sigma)$ terms with $q<n$ in the effective potential is robust.
This is due to the magnetic $\mathbb Z_n$ symmetry of the modified Villain theory. 
This can be seen, for instance, in \eqref{eq:mvillainpartition}, where the symmetry transformation is $\sigma_z \to \sigma_z + 2\pi/n$ and $k_z \to k_z - 1$.

\bibliography{QFT-Mithat-2}

\begin{thebibliography}{32}%
\makeatletter
\providecommand \@ifxundefined [1]{%
 \@ifx{#1\undefined}
}%
\providecommand \@ifnum [1]{%
 \ifnum #1\expandafter \@firstoftwo
 \else \expandafter \@secondoftwo
 \fi
}%
\providecommand \@ifx [1]{%
 \ifx #1\expandafter \@firstoftwo
 \else \expandafter \@secondoftwo
 \fi
}%
\providecommand \natexlab [1]{#1}%
\providecommand \enquote  [1]{``#1''}%
\providecommand \bibnamefont  [1]{#1}%
\providecommand \bibfnamefont [1]{#1}%
\providecommand \citenamefont [1]{#1}%
\providecommand \href@noop [0]{\@secondoftwo}%
\providecommand \href [0]{\begingroup \@sanitize@url \@href}%
\providecommand \@href[1]{\@@startlink{#1}\@@href}%
\providecommand \@@href[1]{\endgroup#1\@@endlink}%
\providecommand \@sanitize@url [0]{\catcode `\\12\catcode `\$12\catcode `\&12\catcode `\#12\catcode `\^12\catcode `\_12\catcode `\%12\relax}%
\providecommand \@@startlink[1]{}%
\providecommand \@@endlink[0]{}%
\providecommand \url  [0]{\begingroup\@sanitize@url \@url }%
\providecommand \@url [1]{\endgroup\@href {#1}{\urlprefix }}%
\providecommand \urlprefix  [0]{URL }%
\providecommand \Eprint [0]{\href }%
\providecommand \doibase [0]{https://doi.org/}%
\providecommand \selectlanguage [0]{\@gobble}%
\providecommand \bibinfo  [0]{\@secondoftwo}%
\providecommand \bibfield  [0]{\@secondoftwo}%
\providecommand \translation [1]{[#1]}%
\providecommand \BibitemOpen [0]{}%
\providecommand \bibitemStop [0]{}%
\providecommand \bibitemNoStop [0]{.\EOS\space}%
\providecommand \EOS [0]{\spacefactor3000\relax}%
\providecommand \BibitemShut  [1]{\csname bibitem#1\endcsname}%
\let\auto@bib@innerbib\@empty
\bibitem [{\citenamefont {Polyakov}(1975)}]{Polyakov:1975rs}%
  \BibitemOpen
  \bibfield  {author} {\bibinfo {author} {\bibfnamefont {A.~M.}\ \bibnamefont {Polyakov}},\ }\href {https://doi.org/10.1016/0370-2693(75)90162-8} {\bibfield  {journal} {\bibinfo  {journal} {Phys. Lett.}\ }\textbf {\bibinfo {volume} {B59}},\ \bibinfo {pages} {82} (\bibinfo {year} {1975})}\BibitemShut {NoStop}%
\bibitem [{\citenamefont {Polyakov}(1987)}]{Polyakov:1987ez}%
  \BibitemOpen
  \bibfield  {author} {\bibinfo {author} {\bibfnamefont {A.}~\bibnamefont {Polyakov}},\ }\href@noop {} {\emph {\bibinfo {title} {Gauge Fields and Strings (Mathematical Reports,)}}},\ \bibinfo {edition} {1st}\ ed.\ (\bibinfo  {publisher} {CRC Press},\ \bibinfo {year} {1987})\BibitemShut {NoStop}%
\bibitem [{\citenamefont {Wen}(2007)}]{Wen200710}%
  \BibitemOpen
  \bibfield  {author} {\bibinfo {author} {\bibfnamefont {X.-G.}\ \bibnamefont {Wen}},\ }\href@noop {} {\emph {\bibinfo {title} {{Quantum Field Theory of Many-body Systems: From the Origin of Sound to an Origin of Light and Electrons (Oxford Graduate Texts)}}}},\ \bibinfo {edition} {reissue}\ ed.\ (\bibinfo  {publisher} {Oxford Univ Pr (Txt)},\ \bibinfo {year} {2007})\BibitemShut {NoStop}%
\bibitem [{\citenamefont {Unsal}(2009)}]{Unsal:2007jx}%
  \BibitemOpen
  \bibfield  {author} {\bibinfo {author} {\bibfnamefont {M.}~\bibnamefont {Unsal}},\ }\href {https://doi.org/10.1103/PhysRevD.80.065001} {\bibfield  {journal} {\bibinfo  {journal} {Phys. Rev.}\ }\textbf {\bibinfo {volume} {D80}},\ \bibinfo {pages} {065001} (\bibinfo {year} {2009})},\ \Eprint {https://arxiv.org/abs/0709.3269} {arXiv:0709.3269 [hep-th]} \BibitemShut {NoStop}%
\bibitem [{\citenamefont {Unsal}(2008)}]{Unsal:2007vu}%
  \BibitemOpen
  \bibfield  {author} {\bibinfo {author} {\bibfnamefont {M.}~\bibnamefont {Unsal}},\ }\href {https://doi.org/10.1103/PhysRevLett.100.032005} {\bibfield  {journal} {\bibinfo  {journal} {Phys. Rev. Lett.}\ }\textbf {\bibinfo {volume} {100}},\ \bibinfo {pages} {032005} (\bibinfo {year} {2008})},\ \Eprint {https://arxiv.org/abs/0708.1772} {arXiv:0708.1772 [hep-th]} \BibitemShut {NoStop}%
\bibitem [{\citenamefont {Unsal}\ and\ \citenamefont {Yaffe}(2008)}]{Unsal:2008ch}%
  \BibitemOpen
  \bibfield  {author} {\bibinfo {author} {\bibfnamefont {M.}~\bibnamefont {Unsal}}\ and\ \bibinfo {author} {\bibfnamefont {L.~G.}\ \bibnamefont {Yaffe}},\ }\href {https://doi.org/10.1103/PhysRevD.78.065035} {\bibfield  {journal} {\bibinfo  {journal} {Phys. Rev.}\ }\textbf {\bibinfo {volume} {D78}},\ \bibinfo {pages} {065035} (\bibinfo {year} {2008})},\ \Eprint {https://arxiv.org/abs/0803.0344} {arXiv:0803.0344 [hep-th]} \BibitemShut {NoStop}%
\bibitem [{\citenamefont {Anber}\ \emph {et~al.}(2015)\citenamefont {Anber}, \citenamefont {Poppitz},\ and\ \citenamefont {Sulejmanpasic}}]{Anber:2015kea}%
  \BibitemOpen
  \bibfield  {author} {\bibinfo {author} {\bibfnamefont {M.~M.}\ \bibnamefont {Anber}}, \bibinfo {author} {\bibfnamefont {E.}~\bibnamefont {Poppitz}},\ and\ \bibinfo {author} {\bibfnamefont {T.}~\bibnamefont {Sulejmanpasic}},\ }\href {https://doi.org/10.1103/PhysRevD.92.021701} {\bibfield  {journal} {\bibinfo  {journal} {Phys. Rev.}\ }\textbf {\bibinfo {volume} {D92}},\ \bibinfo {pages} {021701} (\bibinfo {year} {2015})},\ \Eprint {https://arxiv.org/abs/1501.06773} {arXiv:1501.06773 [hep-th]} \BibitemShut {NoStop}%
\bibitem [{\citenamefont {Gross}\ \emph {et~al.}(1996)\citenamefont {Gross}, \citenamefont {Klebanov}, \citenamefont {Matytsin},\ and\ \citenamefont {Smilga}}]{Gross:1995bp}%
  \BibitemOpen
  \bibfield  {author} {\bibinfo {author} {\bibfnamefont {D.~J.}\ \bibnamefont {Gross}}, \bibinfo {author} {\bibfnamefont {I.~R.}\ \bibnamefont {Klebanov}}, \bibinfo {author} {\bibfnamefont {A.~V.}\ \bibnamefont {Matytsin}},\ and\ \bibinfo {author} {\bibfnamefont {A.~V.}\ \bibnamefont {Smilga}},\ }\href {https://doi.org/10.1016/0550-3213(95)00655-9} {\bibfield  {journal} {\bibinfo  {journal} {Nucl. Phys.}\ }\textbf {\bibinfo {volume} {B461}},\ \bibinfo {pages} {109} (\bibinfo {year} {1996})},\ \Eprint {https://arxiv.org/abs/hep-th/9511104} {arXiv:hep-th/9511104 [hep-th]} \BibitemShut {NoStop}%
\bibitem [{\citenamefont {Cherman}\ \emph {et~al.}(2020)\citenamefont {Cherman}, \citenamefont {Jacobson}, \citenamefont {Tanizaki},\ and\ \citenamefont {\"Unsal}}]{Cherman:2019hbq}%
  \BibitemOpen
  \bibfield  {author} {\bibinfo {author} {\bibfnamefont {A.}~\bibnamefont {Cherman}}, \bibinfo {author} {\bibfnamefont {T.}~\bibnamefont {Jacobson}}, \bibinfo {author} {\bibfnamefont {Y.}~\bibnamefont {Tanizaki}},\ and\ \bibinfo {author} {\bibfnamefont {M.}~\bibnamefont {\"Unsal}},\ }\href {https://doi.org/10.21468/SciPostPhys.8.5.072} {\bibfield  {journal} {\bibinfo  {journal} {SciPost Phys.}\ }\textbf {\bibinfo {volume} {8}},\ \bibinfo {pages} {072} (\bibinfo {year} {2020})},\ \Eprint {https://arxiv.org/abs/1908.09858} {arXiv:1908.09858 [hep-th]} \BibitemShut {NoStop}%
\bibitem [{\citenamefont {Komargodski}\ \emph {et~al.}(2021)\citenamefont {Komargodski}, \citenamefont {Ohmori}, \citenamefont {Roumpedakis},\ and\ \citenamefont {Seifnashri}}]{Komargodski:2020mxz}%
  \BibitemOpen
  \bibfield  {author} {\bibinfo {author} {\bibfnamefont {Z.}~\bibnamefont {Komargodski}}, \bibinfo {author} {\bibfnamefont {K.}~\bibnamefont {Ohmori}}, \bibinfo {author} {\bibfnamefont {K.}~\bibnamefont {Roumpedakis}},\ and\ \bibinfo {author} {\bibfnamefont {S.}~\bibnamefont {Seifnashri}},\ }\href {https://doi.org/10.1007/JHEP03(2021)103} {\bibfield  {journal} {\bibinfo  {journal} {JHEP}\ }\textbf {\bibinfo {volume} {03}},\ \bibinfo {pages} {103}},\ \Eprint {https://arxiv.org/abs/2008.07567} {arXiv:2008.07567 [hep-th]} \BibitemShut {NoStop}%
\bibitem [{\citenamefont {Haldane}(1988)}]{Haldane:1988zz}%
  \BibitemOpen
  \bibfield  {author} {\bibinfo {author} {\bibfnamefont {F.~D.~M.}\ \bibnamefont {Haldane}},\ }\href {https://doi.org/10.1103/PhysRevLett.61.1029} {\bibfield  {journal} {\bibinfo  {journal} {Phys. Rev. Lett.}\ }\textbf {\bibinfo {volume} {61}},\ \bibinfo {pages} {1029} (\bibinfo {year} {1988})}\BibitemShut {NoStop}%
\bibitem [{\citenamefont {Read}\ and\ \citenamefont {Sachdev}(1989)}]{Read:1989zz}%
  \BibitemOpen
  \bibfield  {author} {\bibinfo {author} {\bibfnamefont {N.}~\bibnamefont {Read}}\ and\ \bibinfo {author} {\bibfnamefont {S.}~\bibnamefont {Sachdev}},\ }\href {https://doi.org/10.1103/PhysRevLett.62.1694} {\bibfield  {journal} {\bibinfo  {journal} {Phys. Rev. Lett.}\ }\textbf {\bibinfo {volume} {62}},\ \bibinfo {pages} {1694} (\bibinfo {year} {1989})}\BibitemShut {NoStop}%
\bibitem [{\citenamefont {Read}\ and\ \citenamefont {Sachdev}(1990)}]{Read:1990zza}%
  \BibitemOpen
  \bibfield  {author} {\bibinfo {author} {\bibfnamefont {N.}~\bibnamefont {Read}}\ and\ \bibinfo {author} {\bibfnamefont {S.}~\bibnamefont {Sachdev}},\ }\href {https://doi.org/10.1103/PhysRevB.42.4568} {\bibfield  {journal} {\bibinfo  {journal} {Phys. Rev. B}\ }\textbf {\bibinfo {volume} {42}},\ \bibinfo {pages} {4568} (\bibinfo {year} {1990})}\BibitemShut {NoStop}%
\bibitem [{\citenamefont {Poppitz}\ and\ \citenamefont {Unsal}(2009)}]{Poppitz:2009tw}%
  \BibitemOpen
  \bibfield  {author} {\bibinfo {author} {\bibfnamefont {E.}~\bibnamefont {Poppitz}}\ and\ \bibinfo {author} {\bibfnamefont {M.}~\bibnamefont {Unsal}},\ }\href {https://doi.org/10.1088/1126-6708/2009/12/011} {\bibfield  {journal} {\bibinfo  {journal} {JHEP}\ }\textbf {\bibinfo {volume} {12}},\ \bibinfo {pages} {011}},\ \Eprint {https://arxiv.org/abs/0910.1245} {arXiv:0910.1245 [hep-th]} \BibitemShut {NoStop}%
\bibitem [{\citenamefont {Anber}\ and\ \citenamefont {Poppitz}(2011)}]{Anber:2011de}%
  \BibitemOpen
  \bibfield  {author} {\bibinfo {author} {\bibfnamefont {M.~M.}\ \bibnamefont {Anber}}\ and\ \bibinfo {author} {\bibfnamefont {E.}~\bibnamefont {Poppitz}},\ }\href {https://doi.org/10.1007/JHEP06(2011)136} {\bibfield  {journal} {\bibinfo  {journal} {JHEP}\ }\textbf {\bibinfo {volume} {06}},\ \bibinfo {pages} {136}},\ \Eprint {https://arxiv.org/abs/1105.0940} {arXiv:1105.0940 [hep-th]} \BibitemShut {NoStop}%
\bibitem [{\citenamefont {Sulejmanpasic}\ \emph {et~al.}(2017)\citenamefont {Sulejmanpasic}, \citenamefont {Shao}, \citenamefont {Sandvik},\ and\ \citenamefont {Unsal}}]{Sulejmanpasic:2016uwq}%
  \BibitemOpen
  \bibfield  {author} {\bibinfo {author} {\bibfnamefont {T.}~\bibnamefont {Sulejmanpasic}}, \bibinfo {author} {\bibfnamefont {H.}~\bibnamefont {Shao}}, \bibinfo {author} {\bibfnamefont {A.}~\bibnamefont {Sandvik}},\ and\ \bibinfo {author} {\bibfnamefont {M.}~\bibnamefont {Unsal}},\ }\href {https://doi.org/10.1103/PhysRevLett.119.091601} {\bibfield  {journal} {\bibinfo  {journal} {Phys. Rev. Lett.}\ }\textbf {\bibinfo {volume} {119}},\ \bibinfo {pages} {091601} (\bibinfo {year} {2017})},\ \Eprint {https://arxiv.org/abs/1608.09011} {arXiv:1608.09011 [hep-th]} \BibitemShut {NoStop}%
\bibitem [{\citenamefont {Banerjee}\ \emph {et~al.}(2024)\citenamefont {Banerjee}, \citenamefont {Banerjee}, \citenamefont {Kanwar}, \citenamefont {Mariani}, \citenamefont {Rindlisbacher},\ and\ \citenamefont {Wiese}}]{Banerjee:2023pnb}%
  \BibitemOpen
  \bibfield  {author} {\bibinfo {author} {\bibfnamefont {A.}~\bibnamefont {Banerjee}}, \bibinfo {author} {\bibfnamefont {D.}~\bibnamefont {Banerjee}}, \bibinfo {author} {\bibfnamefont {G.}~\bibnamefont {Kanwar}}, \bibinfo {author} {\bibfnamefont {A.}~\bibnamefont {Mariani}}, \bibinfo {author} {\bibfnamefont {T.}~\bibnamefont {Rindlisbacher}},\ and\ \bibinfo {author} {\bibfnamefont {U.~J.}\ \bibnamefont {Wiese}},\ }\href {https://doi.org/10.1103/PhysRevD.109.014506} {\bibfield  {journal} {\bibinfo  {journal} {Phys. Rev. D}\ }\textbf {\bibinfo {volume} {109}},\ \bibinfo {pages} {014506} (\bibinfo {year} {2024})},\ \Eprint {https://arxiv.org/abs/2309.17109} {arXiv:2309.17109 [hep-lat]} \BibitemShut {NoStop}%
\bibitem [{\citenamefont {Banerjee}\ \emph {et~al.}(2013)\citenamefont {Banerjee}, \citenamefont {Jiang}, \citenamefont {Widmer},\ and\ \citenamefont {Wiese}}]{Banerjee:2013dda}%
  \BibitemOpen
  \bibfield  {author} {\bibinfo {author} {\bibfnamefont {D.}~\bibnamefont {Banerjee}}, \bibinfo {author} {\bibfnamefont {F.~J.}\ \bibnamefont {Jiang}}, \bibinfo {author} {\bibfnamefont {P.}~\bibnamefont {Widmer}},\ and\ \bibinfo {author} {\bibfnamefont {U.~J.}\ \bibnamefont {Wiese}},\ }\href {https://doi.org/10.1088/1742-5468/2013/12/P12010} {\bibfield  {journal} {\bibinfo  {journal} {J. Stat. Mech.}\ }\textbf {\bibinfo {volume} {1312}},\ \bibinfo {pages} {P12010} (\bibinfo {year} {2013})},\ \Eprint {https://arxiv.org/abs/1303.6858} {arXiv:1303.6858 [cond-mat.str-el]} \BibitemShut {NoStop}%
\bibitem [{\citenamefont {Banerjee}\ \emph {et~al.}(2016)\citenamefont {Banerjee}, \citenamefont {B\"ogli}, \citenamefont {Hofmann}, \citenamefont {Jiang}, \citenamefont {Widmer},\ and\ \citenamefont {Wiese}}]{Banerjee:2015pnt}%
  \BibitemOpen
  \bibfield  {author} {\bibinfo {author} {\bibfnamefont {D.}~\bibnamefont {Banerjee}}, \bibinfo {author} {\bibfnamefont {M.}~\bibnamefont {B\"ogli}}, \bibinfo {author} {\bibfnamefont {C.~P.}\ \bibnamefont {Hofmann}}, \bibinfo {author} {\bibfnamefont {F.~J.}\ \bibnamefont {Jiang}}, \bibinfo {author} {\bibfnamefont {P.}~\bibnamefont {Widmer}},\ and\ \bibinfo {author} {\bibfnamefont {U.~J.}\ \bibnamefont {Wiese}},\ }\href {https://doi.org/10.1103/PhysRevB.94.115120} {\bibfield  {journal} {\bibinfo  {journal} {Phys. Rev. B}\ }\textbf {\bibinfo {volume} {94}},\ \bibinfo {pages} {115120} (\bibinfo {year} {2016})},\ \Eprint {https://arxiv.org/abs/1511.00881} {arXiv:1511.00881 [cond-mat.str-el]} \BibitemShut {NoStop}%
\bibitem [{\citenamefont {Banerjee}\ \emph {et~al.}(2014)\citenamefont {Banerjee}, \citenamefont {B\"ogli}, \citenamefont {Hofmann}, \citenamefont {Jiang}, \citenamefont {Widmer},\ and\ \citenamefont {Wiese}}]{Banerjee:2014wpa}%
  \BibitemOpen
  \bibfield  {author} {\bibinfo {author} {\bibfnamefont {D.}~\bibnamefont {Banerjee}}, \bibinfo {author} {\bibfnamefont {M.}~\bibnamefont {B\"ogli}}, \bibinfo {author} {\bibfnamefont {C.~P.}\ \bibnamefont {Hofmann}}, \bibinfo {author} {\bibfnamefont {F.~J.}\ \bibnamefont {Jiang}}, \bibinfo {author} {\bibfnamefont {P.}~\bibnamefont {Widmer}},\ and\ \bibinfo {author} {\bibfnamefont {U.~J.}\ \bibnamefont {Wiese}},\ }\href {https://doi.org/10.1103/PhysRevB.90.245143} {\bibfield  {journal} {\bibinfo  {journal} {Phys. Rev. B}\ }\textbf {\bibinfo {volume} {90}},\ \bibinfo {pages} {245143} (\bibinfo {year} {2014})},\ \Eprint {https://arxiv.org/abs/1406.2077} {arXiv:1406.2077 [cond-mat.str-el]} \BibitemShut {NoStop}%
\bibitem [{\citenamefont {Sulejmanpasic}\ and\ \citenamefont {Gattringer}(2019)}]{Sulejmanpasic:2019ytl}%
  \BibitemOpen
  \bibfield  {author} {\bibinfo {author} {\bibfnamefont {T.}~\bibnamefont {Sulejmanpasic}}\ and\ \bibinfo {author} {\bibfnamefont {C.}~\bibnamefont {Gattringer}},\ }\href {https://doi.org/10.1016/j.nuclphysb.2019.114616} {\bibfield  {journal} {\bibinfo  {journal} {Nucl. Phys.}\ }\textbf {\bibinfo {volume} {B943}},\ \bibinfo {pages} {114616} (\bibinfo {year} {2019})},\ \Eprint {https://arxiv.org/abs/1901.02637} {arXiv:1901.02637 [hep-lat]} \BibitemShut {NoStop}%
\bibitem [{\citenamefont {Hayashi}\ and\ \citenamefont {Tanizaki}(2024)}]{Hayashi:2024yjc}%
  \BibitemOpen
  \bibfield  {author} {\bibinfo {author} {\bibfnamefont {Y.}~\bibnamefont {Hayashi}}\ and\ \bibinfo {author} {\bibfnamefont {Y.}~\bibnamefont {Tanizaki}},\ }\href {https://doi.org/10.1103/PhysRevLett.133.171902} {\bibfield  {journal} {\bibinfo  {journal} {Phys. Rev. Lett.}\ }\textbf {\bibinfo {volume} {133}},\ \bibinfo {pages} {171902} (\bibinfo {year} {2024})},\ \Eprint {https://arxiv.org/abs/2405.12402} {arXiv:2405.12402 [hep-th]} \BibitemShut {NoStop}%
\bibitem [{\citenamefont {Tanizaki}\ and\ \citenamefont {\"Unsal}(2022)}]{Tanizaki:2022ngt}%
  \BibitemOpen
  \bibfield  {author} {\bibinfo {author} {\bibfnamefont {Y.}~\bibnamefont {Tanizaki}}\ and\ \bibinfo {author} {\bibfnamefont {M.}~\bibnamefont {\"Unsal}},\ }\href {https://doi.org/10.1093/ptep/ptac042} {\bibfield  {journal} {\bibinfo  {journal} {PTEP}\ }\textbf {\bibinfo {volume} {2022}},\ \bibinfo {pages} {04A108} (\bibinfo {year} {2022})},\ \Eprint {https://arxiv.org/abs/2201.06166} {arXiv:2201.06166 [hep-th]} \BibitemShut {NoStop}%
\bibitem [{\citenamefont {Dunne}\ \emph {et~al.}(2001)\citenamefont {Dunne}, \citenamefont {Kogan}, \citenamefont {Kovner},\ and\ \citenamefont {Tekin}}]{Dunne:2000vp}%
  \BibitemOpen
  \bibfield  {author} {\bibinfo {author} {\bibfnamefont {G.~V.}\ \bibnamefont {Dunne}}, \bibinfo {author} {\bibfnamefont {I.~I.}\ \bibnamefont {Kogan}}, \bibinfo {author} {\bibfnamefont {A.}~\bibnamefont {Kovner}},\ and\ \bibinfo {author} {\bibfnamefont {B.}~\bibnamefont {Tekin}},\ }\href {https://doi.org/10.1088/1126-6708/2001/01/032} {\bibfield  {journal} {\bibinfo  {journal} {JHEP}\ }\textbf {\bibinfo {volume} {01}},\ \bibinfo {pages} {032}},\ \Eprint {https://arxiv.org/abs/hep-th/0010201} {arXiv:hep-th/0010201} \BibitemShut {NoStop}%
\bibitem [{\citenamefont {Kovchegov}\ and\ \citenamefont {Son}(2003)}]{Kovchegov:2002vi}%
  \BibitemOpen
  \bibfield  {author} {\bibinfo {author} {\bibfnamefont {Y.~V.}\ \bibnamefont {Kovchegov}}\ and\ \bibinfo {author} {\bibfnamefont {D.~T.}\ \bibnamefont {Son}},\ }\href {https://doi.org/10.1088/1126-6708/2003/01/050} {\bibfield  {journal} {\bibinfo  {journal} {JHEP}\ }\textbf {\bibinfo {volume} {01}},\ \bibinfo {pages} {050}},\ \Eprint {https://arxiv.org/abs/hep-th/0212230} {arXiv:hep-th/0212230} \BibitemShut {NoStop}%
\bibitem [{\citenamefont {Simic}\ and\ \citenamefont {Unsal}(2012)}]{Simic:2010sv}%
  \BibitemOpen
  \bibfield  {author} {\bibinfo {author} {\bibfnamefont {D.}~\bibnamefont {Simic}}\ and\ \bibinfo {author} {\bibfnamefont {M.}~\bibnamefont {Unsal}},\ }\href {https://doi.org/10.1103/PhysRevD.85.105027} {\bibfield  {journal} {\bibinfo  {journal} {Phys. Rev. D}\ }\textbf {\bibinfo {volume} {85}},\ \bibinfo {pages} {105027} (\bibinfo {year} {2012})},\ \Eprint {https://arxiv.org/abs/1010.5515} {arXiv:1010.5515 [hep-th]} \BibitemShut {NoStop}%
\bibitem [{\citenamefont {Greensite}(2003)}]{Greensite:2003bk}%
  \BibitemOpen
  \bibfield  {author} {\bibinfo {author} {\bibfnamefont {J.}~\bibnamefont {Greensite}},\ }\href {https://doi.org/10.1016/S0146-6410(03)90012-3} {\bibfield  {journal} {\bibinfo  {journal} {Prog. Part. Nucl. Phys.}\ }\textbf {\bibinfo {volume} {51}},\ \bibinfo {pages} {1} (\bibinfo {year} {2003})},\ \Eprint {https://arxiv.org/abs/hep-lat/0301023} {arXiv:hep-lat/0301023} \BibitemShut {NoStop}%
\bibitem [{\citenamefont {Hayashi}\ \emph {et~al.}(2024)\citenamefont {Hayashi}, \citenamefont {Misumi},\ and\ \citenamefont {Tanizaki}}]{Hayashi:2024psa}%
  \BibitemOpen
  \bibfield  {author} {\bibinfo {author} {\bibfnamefont {Y.}~\bibnamefont {Hayashi}}, \bibinfo {author} {\bibfnamefont {T.}~\bibnamefont {Misumi}},\ and\ \bibinfo {author} {\bibfnamefont {Y.}~\bibnamefont {Tanizaki}},\ }\href@noop {} {\  (\bibinfo {year} {2024})},\ \Eprint {https://arxiv.org/abs/2410.21392} {arXiv:2410.21392 [hep-th]} \BibitemShut {NoStop}%
\bibitem [{\citenamefont {G\"uvendik}\ \emph {et~al.}(2024)\citenamefont {G\"uvendik}, \citenamefont {Schaefer},\ and\ \citenamefont {\"Unsal}}]{Guvendik:2024umd}%
  \BibitemOpen
  \bibfield  {author} {\bibinfo {author} {\bibfnamefont {C.}~\bibnamefont {G\"uvendik}}, \bibinfo {author} {\bibfnamefont {T.}~\bibnamefont {Schaefer}},\ and\ \bibinfo {author} {\bibfnamefont {M.}~\bibnamefont {\"Unsal}},\ }\href@noop {} {\  (\bibinfo {year} {2024})},\ \Eprint {https://arxiv.org/abs/2405.13696} {arXiv:2405.13696 [hep-th]} \BibitemShut {NoStop}%
\bibitem [{\citenamefont {Schwinger}\ \emph {et~al.}(1998)\citenamefont {Schwinger}, \citenamefont {Deraad~Jr.}, \citenamefont {Milton},\ and\ \citenamefont {Tsai}}]{Schwinger}%
  \BibitemOpen
  \bibfield  {author} {\bibinfo {author} {\bibfnamefont {J.}~\bibnamefont {Schwinger}}, \bibinfo {author} {\bibfnamefont {L.}~\bibnamefont {Deraad~Jr.}}, \bibinfo {author} {\bibfnamefont {K.}~\bibnamefont {Milton}},\ and\ \bibinfo {author} {\bibfnamefont {W.-Y.}\ \bibnamefont {Tsai}},\ }\href {https://doi.org/10.1201/9780429503542} {\emph {\bibinfo {title} {{Classical Electrodynamics}}}}\ (\bibinfo  {publisher} {CRC Press},\ \bibinfo {year} {1998})\BibitemShut {NoStop}%
\bibitem [{\citenamefont {Gorantla}\ \emph {et~al.}(2021)\citenamefont {Gorantla}, \citenamefont {Lam}, \citenamefont {Seiberg},\ and\ \citenamefont {Shao}}]{Gorantla:2021svj}%
  \BibitemOpen
  \bibfield  {author} {\bibinfo {author} {\bibfnamefont {P.}~\bibnamefont {Gorantla}}, \bibinfo {author} {\bibfnamefont {H.~T.}\ \bibnamefont {Lam}}, \bibinfo {author} {\bibfnamefont {N.}~\bibnamefont {Seiberg}},\ and\ \bibinfo {author} {\bibfnamefont {S.-H.}\ \bibnamefont {Shao}},\ }\href {https://doi.org/10.1063/5.0060808} {\bibfield  {journal} {\bibinfo  {journal} {J. Math. Phys.}\ }\textbf {\bibinfo {volume} {62}},\ \bibinfo {pages} {102301} (\bibinfo {year} {2021})},\ \Eprint {https://arxiv.org/abs/2103.01257} {arXiv:2103.01257 [cond-mat.str-el]} \BibitemShut {NoStop}%
\bibitem [{\citenamefont {Unsal}(2012)}]{Unsal:2012zj}%
  \BibitemOpen
  \bibfield  {author} {\bibinfo {author} {\bibfnamefont {M.}~\bibnamefont {Unsal}},\ }\href {https://doi.org/10.1103/PhysRevD.86.105012} {\bibfield  {journal} {\bibinfo  {journal} {Phys. Rev. D}\ }\textbf {\bibinfo {volume} {86}},\ \bibinfo {pages} {105012} (\bibinfo {year} {2012})},\ \Eprint {https://arxiv.org/abs/1201.6426} {arXiv:1201.6426 [hep-th]} \BibitemShut {NoStop}%
\end{thebibliography}%

\end{document}